\documentstyle[aps,prl,twocolumn,floats]{revtex}
\input epsf

\begin{document}

\title{Random Walks through the Ensemble: Linking Spectral 
Statistics with Wavefunction Correlations in Disordered Metals}
 
\author{J. T. Chalker$^1$, Igor V. Lerner$^2$, and Robert A. Smith$^2$}  
\address{$^1$Theoretical Physics, University of Oxford, 1 Keble Road,
Oxford OX1 3NP, United Kingdom\\ 
$^2$School of Physics and Space Research, University of 
Birmingham, Edgbaston, Birmingham~B15~2TT, United Kingdom
% }\date{29 March 1996} \maketitle \begin{abstract}
%
\\ \date{\it Submitted to PRL
29 March 1996}\medskip} \author{\small\parbox{14cm}{\small \indent
We use a random walk in the ensemble of impurity configurations to
generate a Brownian motion model for energy levels in disordered
conductors. Treating arc-length along the random walk as fictitous
time, the resulting Langevin equation relates spectral statistics to
eigenfunction correlations.  Solving this equation at energy scales
large compared with the mean level spacing, we obtain the spectral form
factor, and its parametric dependence.
\\[2pt] PACS numbers: 05.45+b, 71.25.-s, 72.15.Rn\\[-6pt] }}
 \maketitle\draft
 % \end{abstract}\draft\pacs{PACS numbers: 05.45+b, 71.25.-s, 72.15.Rn}

\bibliographystyle{simpl1} 
 
Statistical properties of the spectra of finite quantum systems 
have been a focus for research in three successive contexts: 
nuclear physics \cite{Wig}, the semiclassical 
limit of quantum mechanics \cite{Gutb}, 
and studies of mesoscopic conductors \cite{Ef:83,AS}. A unifying idea 
is that 
the energies of individual eigenstates are frequently 
neither calculable nor interesting: instead, the concern should be 
with eigenvalue correlations, which typically are independent of 
many details of the Hamiltonian.

Disordered mesoscopic conductors bring to this field both 
new behaviour and a natural ensemble for a statistical 
description - the 
ensemble of impurity configurations. For weak disorder, new behaviour arises because 
there can be a broad window in time, and hence a corresponding energy 
interval, between the scale, $t_{\text{el}}$, for electron scattering 
from 
impurities and that for diffusion across the system, $t_{\text{erg}} 
\sim L^2/D$ (where $L$ and $D$ are the system size and diffusion constant).
And at the mobility edge, specific, critical spectral statistics are 
expected \cite{ShSh,KLAA}.
The ensemble average
provides the departure point for established perturbative \cite{AS} 
and non-perturbative \cite{Ef:83} calculations of spectral correlations 
in disordered metals.

In this paper we describe an alternative approach, in which the energy 
level distribution is averaged along a random walk through the ensemble. 
There are several precedents for study of levels as a function of 
position in the space of Hamiltonians.
Most recently, a number of authors \cite{para},
in particular Szafer, Simons and Altshuler
\cite{ASz}, have investigated parametric statistics: level 
correlations between different points on a smooth path in this space. 
Earlier, in connection with the semiclassical limit, Pechukas 
\cite{Pechukas} 
used motion along such a path, with coordinate 
$\lambda$, to generate a dynamics for the one-dimensional gas formed 
by levels on the energy axis. And originally, in the context of random 
matrix theory (RMT), Dyson \cite{Dsn2}, employed a random walk through 
the matrix ensemble as the foundation for Brownian dynamics of levels, 
with arc-length, $\tau$, being the fictitious time. Pechukas' and 
Dyson's ideas are linked, as shown schematically
 in Fig.\ 1, by the usual relation \cite{Dsn2} between the 
end-to-end distance and the length of a random walk, $\lambda^2\sim \tau$.

In previous work on level dynamics \cite{Pechukas,Dsn2,Yukawa,Bee3},
eigenfunction correlations have 
played no role, either (in the semiclassical limit) by assumption, 
or (for random matrices) by construction,
RMT having 
no preferred basis. In contrast, for disordered metals, the basis of 
position states is singled out: it is in this basis that the impurity 
potential is a diagonal operator. Further, specific eigenfunction 
correlations must be present if, for example, wavepackets spread 
diffusively. We show here that these correlations result in a novel 
Brownian level dynamics, from which we obtain the features of 
spectral correlations particular to mesoscopic conductors.
%%%%%HERE IS THE FIGURE
\begin{figure}
\vspace*{.23cm}
\hspace*{0.05\columnwidth}\epsfxsize= 0.9 \columnwidth
\epsffile{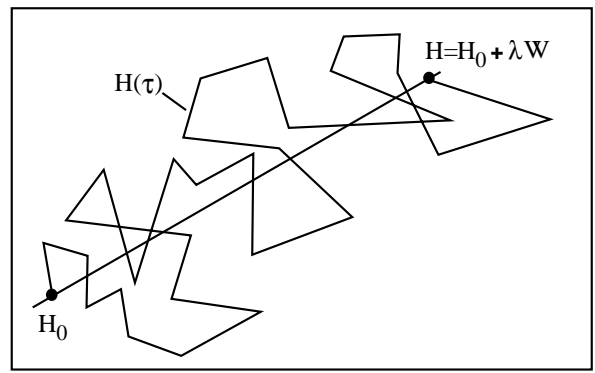}
\vspace*{.20cm}
\caption{Smooth [$H(\lambda)$] and Brownian [$H(\tau)]$ paths through 
the space of Hamiltonians.}
\end{figure}

Our results are expressed in terms of the 
{\it quantum} return probability, $p(t)$, 
which is required as input in this approach. 
Consider the time-evolution 
of a wavepacket that initially occupies a small volume  $l_0^{d}$ 
\cite{bmcom5}: $p(t)$ is the probability density to remain in this 
volume at the time $t$. For weak disorder, the wavepacket spreads 
diffusively,
and $p(t)$ is known from semiclassical arguments; at the 
metal-insulator
transition, scaling ideas relate $p(t)$ to fractal properties
of eigenstates.

We shall be interested in the two-level correlation function (TLCF) 
and the spectral form factor. 
Let $E_n(\tau)$ be the energy levels of a one-parameter family 
of Hamiltonians, $H(\tau)$. The density of
states per unit volume (DoS) is
$\rho(E,\tau) = L^{-d}\sum_n \delta\bigl(E-E_n(\tau)\bigr)\,$,
and the mean level spacing is
$\Delta \!=\! (\rho L^d)^{-1}$, 
with $\rho \!\equiv\!\langle\rho(E,\tau)\rangle$.
The TLCF is 
\begin{equation}
\label{R(E)}
R(E,\tau) = \rho^{-2} \bigl< \rho(E,\tau) \rho(0,0) \bigr>-1\,,
\end{equation}
and the spectral form factor is
\begin{equation}
\label{K(t)}
K(t,\tau) = {1 \over 2 \pi \hbar} \int_{-\infty}^{\infty}
e^{-i E t/\hbar} R(E,\tau)\, dE\,.
\end{equation}
 
We obtain $K(t,\tau)$ 
for times shorter than the Heisenberg time,
$t_{\!_H} \equiv \hbar/\Delta$,  in terms of $p(t)$.
We defer discussion of the parametric dependence until the end of this paper.
The non-parametric form factor, 
\mbox{$K(t)\equiv K(t, \tau=0)$,} is
\begin{eqnarray}
\label{K1}
K(t)=
{|t|\,p(t)\over {1+
(\pi\hbar\rho )^{-1}
 \int_{0^+}^{|t|}p(t') d t'}}
\left({\Delta \over 2\pi\hbar}\right)^{\!\!2}\!\!L^d\,.
\end{eqnarray}
In the diffusive regime, $t_{\text{el}}\alt t\alt t_{\text{erg}}$, 
 $p(t)$ at leading order reduces to the {\it 
classical } return probability for random walks, multiplied by a 
symmetry factor $2/\beta$, where  $\beta=1,\,2 \,\text{or}\, 4$ is 
the usual index
 \cite{RMT}: 
\begin{eqnarray}
\label{dif}
p_{0}(t)  = {2 \over \beta (4 \pi D t)^{d/2} }\,.
\end{eqnarray}
There exist quantum (weak-localization) corrections to 
Eq.\ (\ref{dif}), 
which are smaller by a factor $g_0^{-1}$,
where $g_0\gg1$ is the dimensionless conductance at scale 
of the elastic mean free path, $\ell$. 
Noting that in the ballistic regime,  $t\alt t_{\text{el}}$, $p(t)$
saturates at $p_{0}(t_{\text{el}})\sim 1/\ell^{d}$,
one sees that the second term in the denominator of Eq.\ (\ref{K1}) also 
gives weak localization corrections: it is of order
$(t_{\text{el}}\Delta/\hbar) (L/\ell)^d\sim g_0^{-1}$
for $d>2$, and of order $ g_0^{-1}\ln (t/t_{\text{el}})$ for  $d=2$.  
Neglecting  these corrections in both the numerator and denominator of 
Eq.\ (\ref{K1}), it reduces to   
\begin{eqnarray}
\label{K3}
K_0(t)&=&
\left({\Delta \over 2\pi\hbar}\right)^{\!\!2}\!\!L^d\,
|t|\,p_{0}(t)\,.
\end{eqnarray}
To leading order, this expression is also valid in
 the ergodic regime, $t_{\text{erg}}\alt t\ll t_{\!_H}$, where
the classical return probability saturates at $(2/\beta) L^d$ so that 
the
second term in the denominator in
Eq.\ (\ref{K1}) is of order $t/t_{\!_H}\ll1$  and  may be neglected.

Equation (\ref{K3})
 coincides with the result obtained in Ref.\ \cite{ImSm},
using the diagonal approximation in semiclassical periodic-orbit
theory.
The Fourier transform of this expression corresponds to 
 the TLCF obtained originally by Altshuler and Shklovskii from 
the lowest order of the diagrammatic expansion\cite{AS}.
In the diffusive regime, $R(s,0) \sim A_d \, s^{d/2-2}$,
and in the ergodic regime, $R(s,0) \sim -1/s^2$ \cite{bmcom2}, 
where $s\!\equiv\!E/\Delta$, and $A_d$ is a numerical coefficient which is zero
for $d\!=\!2$\cite{KL:94}.

There are indications that our approach is useful beyond the diagonal
approximation. We have checked  Eq.\ (\ref{K1}) in the diffusive regime
 by direct diagrammatic expansion of both sides in powers of
$g_0^{-1}$, calculating the first two non-trivial orders for $d=2$.
This is the most interesting case because, for $d=2$,
$p_{0}(t)\propto 1/t$, so that $K(t)$ is constant and $R(s,0)=0$
in the diagonal approximation.
We find \cite{future} that
Eq.\ (\ref{K1}) is {\it exact} up to  order  $g_0^{-2}$. 
For $\beta\!=\!1,\,4$ this means that both the leading ($g_0^{-1}$)
contribution to $R(s,0)$  
\cite{KL:94} and the $g_0^{-2}$ correction are exact. 
In the unitary case ($\beta\!=\!2$) the first order 
terms in the numerator and denominator of Eq.\ (\ref{K1})
cancel, in agreement with the known absence of
corrections to $K_0(t)$ at this order, and it is the $g_0^{-2}$ contribution
which governs  $R(s,0)$; again it is given exactly by  Eq.\ (\ref{K1}). 
For $d>2$, first order corrections to the diagonal approximation
are small by an additional factor, $(t_{\text{el}}/t)^{d/2-1}$;
Eq.\ (\ref{K1}) reproduces their form but not the numerical 
coefficient.
On the other hand, at the mobility edge in $d>2$, Eq.\ (\ref{K1}) 
is consistent with the results of independent calculations, as we 
discuss at the end of the paper.

We next turn to the derivation of our results. 
Let the fictitious time
$\tau$ parameterize a Brownian path through an ensemble
of disordered conductors, so that 
\begin{mathletters}
\label{LangP}
\begin{eqnarray}
\label{LangEns}
H(\tau)&=&H_0+\int^{\tau}_0 \!\!\! d\tau'\, V(\tau',\bbox{r})\,,\\
\label{AndHam}
H_0 &=& -{\hbar^2\over 2m}\nabla^2 + U(\bbox{r})\,.
\end{eqnarray}
\end{mathletters}
We take $U(\bbox{r})$ and $V(\tau,\bbox{r})$ to be 
Gaussian distributed with zero average and 
\begin{eqnarray}
\label{VV}
\begin{array}{rclrcl}
\bigl< U(\bbox{r})U(\bbox{r}')\bigr>&\!\!
=\!\!&(\hbar/2\pi\rho t_{\text{el}})\delta(\bbox{r}\!-\!\bbox{r}')\,,\\[6pt]
\bigl< V(\tau ,\bbox{r})V(\tau' ,\bbox{r}')\bigr>&
\!\!=\!\!&v^2L^d\,\delta(\tau-\tau')\,\delta(\bbox{r}\!-\!\bbox{r}' )\,.
\end{array}
\end{eqnarray}

We use perturbation theory to second order to calculate the change 
of $E_n(\tau)$ in response to $V(\tau,{\bbox{r}}) \delta\tau$:
\begin{eqnarray*}
\delta E_n=V_{nn}\,+\sum_{m\ne n}{V_{mn}V_{nm}\over
E_n-E_m}\,,\\
V_{nm}\equiv\int^{\tau+\delta \tau}_{\tau}
\!\!\!d\tau'\,\left<n\,|V(\tau',\bbox{r})|\,m\right>\,,
\end{eqnarray*}
where $\langle\bbox{r}\,|\,n\rangle\equiv\psi_n(\tau',\bbox{r})$ are
the corresponding eigenfunctions of $H(\tau')$. We average
over $V$ noting that 
\begin{mathletters}
\label{NCorr}
\begin{eqnarray}
\label{NewCorr}
\bigl< V_{nm}V_{mn}\bigr>=
\bigl< V_{nn}V_{mm}\bigr>=v^2 \int \!\! d\tau \,c_{nm}(\tau)\,,\\
\label{c}
c_{nm}(\tau)=L^d\int \!d^dr\, |\psi_n(\tau,r)|^2 |\psi_m(\tau,r)|^2.
\end{eqnarray}
\end{mathletters}
The level dynamics thus obeys the Langevin 
equation
\begin{mathletters}
\label{Lang}
\begin{eqnarray}
\label{LangEqn}
\frac{dE_n(\tau)}{d\tau}=v^2\sum_{l\ne 0}
\frac{c_{n,n+l}(\tau)}{E_n(\tau)-E_{n+l}(\tau)} + \xi_n(\tau)\,,
\end{eqnarray}
where $\xi$ is a random force with 
$\left<\xi\right>=0$ and 
\begin{eqnarray}
\label{EtaDef}
\bigl<\xi_n(\tau)\xi_m(\tau')\bigr>=
v^2 \,\delta(\tau-\tau')\,c_{nm}(\tau)\,.
\end{eqnarray}
\end{mathletters}
Both the spectrum of the random force and
 the drift term are expressed in terms of
$c_{nm}(\tau)$, the eigenstate correlation function,
which we later show to be proportional to
the return probability $p(t)$.

The appearance of the eigenstate correlation function in Eqs.\ 
(\ref{NCorr}) is the 
essential difference between our Brownian level dynamics for disordered
conductors, and that of Dyson for random matrices. 
The analogues to Eqs.\ (\ref{NCorr}) in Dyson's work \cite{Dsn2} have 
the basis-independent form 
\begin{eqnarray}
\label{RmtCorr}
\bigl< V_{nm}V_{mn}\bigr>={2}/{\beta}\,,\qquad
\bigl< V_{nn}V_{mm}\bigr>=\delta_{mn}\,.
\end{eqnarray}
In contrast to Eqs.\ (\ref{NCorr}), these display two simplifying 
features:
they are independent of the eigenvectors, and they lead to diagonal 
random
force correlations. As a result, the limiting level distribution in RMT 
can
be calculated exactly, proceeding via a Fokker-Planck formulation 
\cite{Dsn2}.
For disordered conductors, the Langevin equation for energy levels 
is not
closed, since Eq.\ (\ref{LangEqn}) 
involves
$c_{nm}(\tau)$, 
and the random
forces have off-diagonal correlations. In consequence, 
approximations 
are required, which are most transparent within the Langevin 
description.

Two central assumptions are necessary in order to make progress.
We believe that they are reasonable provided one considers
only behaviour at energy scales much larger than the mean level spacing.
 First, we replace $c_{nm}(\tau)$ by its 
average over the ensemble of $H_0$,
which, within the energy window of interest, is a function only of the energy
difference, $E=E_n-E_m$:
\begin{equation}
\label{c(E)}
\bigl< c_{n,n+l}(\tau) \bigr> \equiv c(E)\,.
\end{equation}
Thus we neglect correlations between fluctuations in $c_{nm}(\tau)$ 
and those in $\rho (E)$ \cite{footnote2}.
Second, we linearize Eqs.\ (\ref{Lang})
 in the deviation of level
separation $E_n-E_m$ from its mean value $(n-m)\Delta$,
\begin{eqnarray*}
E_n(\tau)=n\Delta +\varepsilon_n(\tau)\,.
\end{eqnarray*}
Then  Eq.\ (\ref{LangEqn}) becomes
\begin{equation}
\label{xdot}
{d\varepsilon_n(\tau) \over d\tau} = - v^2 {\sum_{l \not= 0}}
(\varepsilon_{n+l}-\varepsilon_{n}) f(l\Delta) + \xi_n(\tau)
\end{equation}
where  $f(E) \equiv d\varphi(E)/dE$ with $\varphi(E) = c(E)/E$.
This is diagonalized by
Fourier transform,  using as coordinates
\begin{mathletters}
\label{X}
\begin{equation}
{\cal {E}}(t,\tau) = {\Delta \over 2 \pi \hbar} \sum_n 
\varepsilon_n(\tau)
e^{-i\Delta n t/ \hbar}\,,
\end{equation}
which, for  $0<t\ll \hbar/\Delta$, and $\varepsilon_n t/\hbar$
are essentially Fourier components of the DoS:
\begin{equation}
\label{X-rho}
{\cal {E}}(t,\tau) \simeq {i \over 2 \pi\rho t} \int_{-\infty}^{\infty}
\rho(E,\tau) e^{-i E t/\hbar} \, dE\,.
\end{equation}
\end{mathletters}
In these variables, Eqs.\ (\ref{Lang}) become
\begin{mathletters}
\label{Lang2}
\begin{eqnarray}
{d{\cal {E}}(t,\tau)\over d\tau}  = -{2 \pi \hbar v^2\over \Delta}
 F(t){\cal {E}}(t,\tau) + \Xi(t,\tau)\,,\\
\label{eta(t)}
\bigl< \Xi(t,\tau)\, \Xi(t',\tau') \bigr> = 
v^2
\delta(t\!+\!t') \delta(\tau \!-\! \tau') C(t)\,,
\end{eqnarray}
\end{mathletters}
where $C(t)$, $F(t)$ and $\Xi(t,\tau)$ are the Fourier transforms of 
$c(E)$, $f(E)$ and $\xi_n(\tau)$,
defined 
as in Eqs.\ (\ref{X}). For example,
\begin{eqnarray*}
C(t)=\frac{\Delta}{2\pi\hbar} 
\sum_l c(l\Delta)e^{-i\Delta l t/ \hbar}\simeq\int_{-\infty}^{\infty}\!\!\!\!
c(E)e^{-iEt/\hbar}\frac{dE}{2\pi\hbar} \,.
\end{eqnarray*}
It follows from the definition of $f(E)$, after Eq.\ (\ref{xdot}),
 that $F(t)$ is related to $C(t)$. 
Simple manipulations lead to
$
F(t)=(t/\hbar^2) \int_{0}^t C(t') dt'\,,
$
where we have used the fact that $F(t)$ is an even
function of $t$ to fix the constant of integration.
Finally, we note that, since $c(E)=1$ for $|E|\agt \hbar/t_{\text{el}}$,
$C(t)$ contains a $\delta$-function at $t=0$,
 so that
\begin{eqnarray}
\label{F(t)}
F(t)= {|t| \over 2 \hbar^2  } + {|t| \over \hbar^2} 
\int_{0^+}^{|t|} C(t') d
t'\,.\end{eqnarray}

Now we solve Eqs.\ (\ref{Lang2}), fixing  at this point the units of
$\tau$ by setting $v^2 = \Delta  / 2 \pi \hbar$:
\begin{eqnarray*}
\langle {\cal {E}}(t,\tau \!+\! \tau'){\cal {E}}(t',\tau') \rangle =
\delta(t\!+\!t') {C(t)\Delta  \over 4 \pi \hbar F(t)}
e^{-F(t)|\tau|}
\end{eqnarray*}
As ${\cal {E}}(t,\tau )$ is the Fourier transform of $\rho(E,\tau)$, we 
immediately obtain the spectral form factor:
\begin{equation}
\label{K}
K(t,\tau) = 
\left({t \over \hbar} \right)^{\!\! 2} {C(t) \Delta \over 4
\pi \hbar F(t)} e^{-F(t)|\tau|}\,.
\end{equation}
 Note that the form factor is expressed in terms of eigenfunction 
correlations via the interplay of the restoring
force, $F$, and the noise correlator, $C$. It is the latter which is 
responsible for the difference between spectral statistics in the 
ergodic and diffusive regimes.

We relate $F$ and $C$ to
the return probability of a diffusing electron, by 
considering a wavepacket made from the eigenstates of
$H(\tau)$ and concentrated initially near the origin,
in a volume of size $\ell^d$:
\begin{eqnarray*}
\Psi({\bbox{r}},t) =A \sum_n \psi_n({\bbox{0}})^*
\psi_n({\bbox{r}}) e^{-iE_n t / \hbar}\,.
\end{eqnarray*}
Here the summation is limited to ${\cal N}$ levels with energies
$E_n\alt  \hbar/ t_{\text{el}}$
 so that ${\cal N}\sim \hbar/ t_{\text{el}} \Delta$, 
and the normalization constant is $A^2=L^d {\cal N}^{-1}$.
The ensemble-averaged return probability,
$p(t)  =\left<|\Psi({\bbox{0}},t)|^2\right>$,  is given for $t>0$ by
\begin{eqnarray}
 p(t) = A^2 \sum_{nm}\left<
c_{nm} e^{-i(E_n-E_m)t/\hbar}\right>
  = {2 \pi \hbar \rho}\, C(t),
\label{p(t)}
\end{eqnarray}
where we have again (as in the
solution of the linearized Langevin equation) neglected
correlations between the fluctuations of
$\{\varepsilon_n(\tau)\}$ and $c_{nm}(\tau)$.

Substituting for $C(t)$ 
in Eq.\ \ref{K},
we find:
\begin{mathletters}
\label{KM}
\begin{eqnarray}
\label{K2}
K(t,\tau)&=&
\left({\Delta \over 2\pi\hbar}\right)^{\!\!2}\!\!L^d\,
{|t|\,p(t)\over {M(t) }}
\exp \!\left[{-{M(t) \over 2 \hbar^2  }\,|t\,\tau|}\right] \,,\\
\label{M2}
M(t)&=&1 \!+\! {1 \over\pi\hbar\rho }\!
 \int_{0^+}^{|t|}\!\! p(t') d t'\,.
\end{eqnarray}
\end{mathletters}
These expressions, which are the
central result of this work, relate spectral 
statistics to the return probability and thus, via Eq.\ (\ref{p(t)}),
to the wavefunction correlations (\ref{c}). 
For the non-parametric problem, Eq.\ (\ref{K2}) reduces to 
Eq.\ (\ref{K1}), and the ensuing behaviour in the ergodic and 
diffusive regimes has been discussed following Eq.\ (\ref{dif}).

Parametric correlations in these regimes, 
from Eqs.\ (\ref{KM}), are as follows. At leading order
we ignore weak-localization contributions to $p(t)$ and $M(t)$, 
approximating them by $p_0(t)$ and 1, respectively. Then, for $t\ll t_{\!_H}$,
\begin{eqnarray}
\label{Kp}
K(t,\tau)&=&K_0(t)\,e^{-|t\,\tau|/2\hbar^2 }\,.
\end{eqnarray}
Its Fourier transform gives the parametric TLCF in both the ergodic regime
(where it reproduces the envelope of the TLCF 
found in Refs.\ \cite{ASz,Bee3}) and the diffusive regime
(where it is  consistent with the results of diagrammatic calculations). 
For $E=0$ we obtain
\begin{eqnarray}
\label{R2}
R(0,\lambda)&=&
{2\over\beta}\left\{
\begin{array}{ll}
\lambda^{-4},\;&\text{ergodic regime}\\
{B_d g^{-d/2}} \lambda^{d-4},&\text{diffusive regime}
\end{array}\right.
\end{eqnarray}
where $\lambda^2\equiv (\pi\tau/\hbar\Delta)$ and
$B_d$ is a constant which is non-zero in 
$d=2$, so that the leading approximation
is sufficient for the parametric TLCF in $d=2$,
in contrast to the non-parametric TLCF, $R(E,\tau\!=\!0)$.

Finally, we turn to spectral correlations for a system at the 
metal-insulator transition \cite{KLAA,ShSh}. At the critical point, 
power-law decay of the return probability is expected for 
$t_{\text{el}} \ll t \ll t_{\text{erg}} \sim t_{\!_H}$, with 
the power related to a multifractal 
dimension of wavefunctions \cite{eta}:
\begin{eqnarray}
\label{p.crit}
p(t) \propto t^{-1 +\eta/d }\,,
\end{eqnarray}
where $0\leq \eta \equiv d-d_2< d$, and $d_2$ is the dimension associated
with $|\psi|^4$ \cite{footnote3}. 
In consequence, for the same range of $t$, $K(t)$ is constant,
a result obtained previously from a diagrammatic analysis \cite{KLAA}.
As in $d=2$,
the behaviour of $R(E,\tau\!=\!0)$ at the mobility edge is, 
in fact determined \cite{KLAA}, not by this limiting behaviour 
for $K(t)$, but by the way in which the limit is approached 
as $t/t_{\!_H} \to 0$; we defer further discussion to a 
future publication.
{\it Parametric} correlations at the mobility edge have not 
previously been investigated. We find $K(t,\tau) \sim K_0 
\exp(-a\tau |t|^{1+\eta/d})$, where $a$ is a constant, giving
\begin{eqnarray}
\label{crit.para}
R(0,\lambda) \sim  \lambda^{-{2\over 1+ \eta/d}}.
\end{eqnarray}
Thus the multifractal properties of critical eigenstates, via the 
exponent $\eta$,
are reflected in parametric spectral statistics at the mobility edge.

In summary, we have developed a simple and transparent approach for 
calculating spectral correlations in disordered metals. By averaging
along a random walk through the ensemble of impurity configurations, 
we have expressed level statistics in 
all regimes
in terms of the quantum return probability for a spreading wavepacket.

We thank V.~E.~Kravtsov and B.~D.~Simons for numerous 
useful discussions. Support from EPSRC grants GR/GO 2727 (J.T.C.) 
and GR/J35238 (I.V.L. and R.A.S.) is gratefully acknowledged.

%\bibliography{my5,mine1,com} \end{document}

\begin{references}
\bibitem{bmcom5}
It is convenient to choose $l_0 \sim \ell$, the mean
  free path.
\bibitem{bmcom2}
In the ergodic regime, this expression gives only the envelope 
of the TLCF as
  the oscillations in $R(s)$ \cite{RMT} are due to singularities 
in $K(t)$ near
  $t_{\!_H}$, which cannot be accounted for semi-classically.
\bibitem{footnote2} 
At this point, we exclude from consideration the Anderson insulator, 
where one must identify not only the energies of 
eigenstates but also their positions in real space.
\bibitem{footnote3}
No other multifractal dimension enters because, for $t\ll t_{\!_H}$, we have
expressed $K(t, \tau)$ solely in terms of $p(t)$.
\end{references}

\end{document}